\RequirePackage[hyphens]{url}
\documentclass[aps,pra,twocolumn,showpacs,bibnotes,reprint,superscriptaddress,floatfix]{revtex4-1}

\usepackage{hyperref,url}
\usepackage{amsmath}
\usepackage{amsfonts}
\usepackage[capitalise]{cleveref}
\hypersetup{hyperindex,breaklinks,colorlinks=true,linkcolor=blue,citecolor=blue,urlcolor=blue,filecolor=blue}
\usepackage{graphicx}
\usepackage{dcolumn}
\usepackage{verbatim}
\usepackage{bbold}

\newcommand{\bi}{\mathbf{i}}
\newcommand{\bj}{\mathbf{j}}

\begin{document}

\title{Accurate exchange-correlation energies for the warm dense electron gas}

\author{Fionn D. Malone}
\email[Author to whom correspondence should be addressed. Electronic mail:]{f.malone13@imperial.ac.uk}
\affiliation{Department of Physics, Imperial College London, Exhibition Road, London SW7 2AZ, United Kingdom.}
\author{N. S. Blunt}
\affiliation{University Chemical Laboratory, Cambridge University, Lensfield Road, Cambridge CB2 1EW, United Kingdom.}
\affiliation{Max Planck Institute for Solid State Research, Heisenbergstrasse 1, Stuttgart 70569, Germany.}
\author{Ethan W. Brown}
\affiliation{Institute for Theoretical Physics, ETH Z{\"u}rich, Wolfgang Pauli Strasse 27, 8093 Z{\"u}rich, Switzerland.}
\author{\mbox{D.K.K.\  Lee}}
\affiliation{Department of Physics, Imperial College London, Exhibition Road, London SW7 2AZ, United Kingdom.}
\author{J.S. Spencer}
\affiliation{Department of Physics, Imperial College London, Exhibition Road, London SW7 2AZ, United Kingdom.}
\affiliation{Department of Materials, Imperial College London, Exhibition Road, London SW7 2AZ, United Kingdom.}
\author{W.M.C. Foulkes}
\affiliation{Department of Physics, Imperial College London, Exhibition Road, London SW7 2AZ, United Kingdom.}
\author{James J. Shepherd}
\affiliation{Department of Chemistry, Massachusetts Institute of Technology, Cambridge, MA, 02139, United States of America.}
\affiliation{Department of Physics, Imperial College London, Exhibition Road, London SW7 2AZ, United Kingdom.}
\begin{abstract}
Density matrix quantum Monte Carlo (DMQMC) is used to sample exact-on-average $N$-body density matrices for uniform electron gas systems of up to 10$^{124}$ matrix elements via a stochastic solution of the Bloch equation.
The results of these calculations resolve a current debate over the accuracy of the data used to parametrize finite-temperature density functionals.
Exchange-correlation energies calculated using the real-space restricted path-integral formalism and the $k$-space configuration path-integral formalism disagree by up to $\sim$$10$\% at certain reduced temperatures $T/T_F \le 0.5$ and densities $r_s \le 1$.
Our calculations confirm the accuracy of the configuration path-integral Monte Carlo results available at high density and bridge the gap to lower densities, providing trustworthy data in the regime typical of planetary interiors and solids subject to laser irradiation. We demonstrate that DMQMC can calculate free energies directly and present exact free energies for $T/T_F \ge 1$ and $r_s \le 2$.
\end{abstract}

\date{\today}
\maketitle

The uniform electron gas is perhaps the most fundamental model in condensed matter physics.
Core concepts such as Fermi liquid theory \citep{landau1957theory}, quasiparticles and collective excitations \citep{PhysRev.85.338,PhysRev.92.609}, screening \citep{giuliani2005quantum}, the BCS theory of superconductivity \citep{PhysRev.108.1175}, and Hohenberg-Kohn-Sham DFT \citep{PhysRev.136.B864,PhysRev.140.A1133}, were all built on our understanding of the electron gas at low temperature.
A growing interest in matter at extreme conditions, especially in the warm dense regime \citep{Fortney2009} found in inertial confinement fusion experiments \citep{PhysRevB.84.224109}, planetary interiors \citep{Fortney2009} and laser-irradiated solids \citep{Ernstorfer20022009}, has sparked efforts to extend this understanding to much higher temperatures.
This Letter concerns the properties of the electron gas at temperatures comparable to the Fermi energy.

The quantitative successes of ground-state DFT rest on parametrizations of the correlation energy of the electron gas at zero temperature \citep{doi:10.1139/p80-159,PhysRevB.23.5048,PhysRevB.45.13244}.
Errors of a few percent in the correlation functional have large effects on chemical bonding and phase diagrams, so these parametrizations are based on accurate QMC data \citep{PhysRevLett.45.566}.
Thermal DFT \citep{PhysRev.137.A1441} treats thermal, quantum mechanical, many-body, and
material effects explicitly and has emerged as a viable tool
\citep{:/content/aip/journal/pop/23/4/10.1063/1.4947212} for the study of warm dense
matter, but requires as input a similarly accurate parametrization of the
exchange-correlation free energy in the entire temperature-density plane
\citep{PhysRev.137.A1441,brown2014path,PhysRevE.93.063207}.

A significant step towards providing this much needed data was recently made by Brown \emph{et al.}\ \citep{PhysRevLett.110.146405} using the restricted path-integral Monte Carlo method, with local density parametrizations quickly following \citep{PhysRevB.88.081102,PhysRevB.88.115123,PhysRevLett.112.076403}.
Soon after this, however, an alternative technique, configuration PIMC, was applied to the same problem and gave substantially different results \citep{CTPP:CTPP201100012,CTPP:CTPP201400072}.

This Letter resolves the disagreement between the two path-integral methods.
We describe an alternative approach called DMQMC \citep{PhysRevB.89.245124,:/content/aip/journal/jcp/143/4/10.1063/1.4927434}, which is, in principle, exact in a given basis set at any temperature or density.
By introducing a systematically improvable approximation analogous to the initiator approximation of full configuration-interaction quantum Monte Carlo (FCIQMC) \citep{Booth2009,:/content/aip/journal/jcp/132/4/10.1063/1.3302277}, we show that DMQMC can be made capable of treating system sizes comparable to those tackled using path integral methods.
We then use initiator DMQMC to settle the controversy and provide new data that can be corrected to the thermodynamic limit using existing techniques \citep{PhysRevLett.110.146405,PhysRevLett.112.076403}.
Our results are particularly useful at densities above $r_s=1$, where no configuration PIMC results exist and the restricted PIMC results are inaccurate.
Such densities are found in laser-irradiated solids and many planetary interiors.
Finally, we show that exchange-correlation free energies are straightforward to estimate in DMQMC and present directly calculated free-energy data for the electron gas.

{\bf \emph{Warm dense electrons.}--} The electron gas can be described by the dimensionless density parameter $r_s = \tilde{r}_s/a_0$, where $a_0$ is the Bohr radius and $\tilde{r}_s$ is the Wigner-Seitz radius, and by the dimensionless temperature $\Theta=T/T_F$, with $T_F$ the Fermi temperature of a three-dimensional free electron gas of the same density.
In the warm dense regime, $r_s \approx \Theta \approx 1$, perturbative methods \citep{Perrot1984,kraeft1986quantum,PhysRevE.69.046407} fail due to the lack of any small coupling parameter, and numerical techniques such as QMC are required.

Real-space restricted PIMC is regarded as the state-of-the-art method for simulating thermal effects in materials \citep{PhysRevLett.115.176403}.
Very recently, however, Schoof \emph{et al.}\ performed highly accurate simulations using the $k$-space configuration PIMC method \citep{CTPP:CTPP201100012,CTPP:CTPP201400072} and obtained results in substantial disagreement with restricted PIMC for the internal energy of the spin-polarized electron gas in the high-density, low-temperature regime \citep{PhysRevLett.115.130402}.
The same group has also reported disagreements with restricted PIMC at higher temperatures and lower densities, this time by comparing with the direct \citep{PhysRevE.91.033108} and permutation-blocking \citep{:/content/aip/journal/jcp/143/20/10.1063/1.4936145} path-integral approaches.
Groth, Dornheim and coworkers \cite{PhysRevB.93.085102,PhysRevB.93.205134} have applied these methods to the polarized and unpolarized electron gas across the entire density range for temperatures above $\Theta=0.5$, finding better agreement with restricted PIMC for the unpolarized than for the polarized case.
Nevertheless, these disagreements will have to be resolved before finite-temperature DFT can be used with the same degree of confidence as its ground-state counterpart.

{\bf \emph{The DMQMC method.--}}
The exact equilibrium properties of any quantum system can be derived from the canonical $N$-particle density matrix $\hat{\rho}(\beta) = e^{-\beta \hat{H}}$.
A deterministic evaluation of $\hat{\rho}$ is intractable for all but the smallest systems; the storage requirements alone rapidly overwhelm even the most modern of computers.
Instead, we seek a stochastic approach.

The density matrix $\hat{\rho}(\beta)$ obeys the
Bloch equation:
\begin{equation}
    \frac{d \hat{\rho}}{d \beta} = -\frac{1}{2}\{\hat{H}, \hat{\rho}\},\label{eq:bloch}
\end{equation}
where $\{\cdot,\cdot\}$ is the usual anti-commutator.
The form of \cref{eq:bloch} is reminiscent of the imaginary-time Schr{\"o}dinger equation, which may be solved stochastically using projector QMC techniques such as diffusion Monte Carlo \citep{RevModPhys.73.33} and, more recently, FCIQMC \citep{Booth2009}.
Taking inspiration from the FCIQMC method, we solve \cref{eq:bloch} using a collection of signed walkers to represent the elements $\rho_{\bi\bj}=\langle D_\bi|\hat{\rho}|D_\bj\rangle$ of the density matrix, where $|D_\bi\rangle$ is a Slater determinant of plane waves.
The DMQMC algorithm ensures that the expected number of walkers found on density matrix element $\rho_{{\bi}{\bj}}$ at imaginary time $\beta$ is proportional to $\rho_{\bi\bj}(\beta)$.
A simulation proceeds by evolving an initial distribution of walkers at $\beta=0$, chosen to provide an unbiased statistical sample of the initial density matrix ($\rho_{\bi\bj}(\beta=0) = \delta_{\bi\bj}$), using a population dynamics algorithm derived from a simple Euler approximation to \cref{eq:bloch} \citep{PhysRevB.89.245124,:/content/aip/journal/jcp/143/4/10.1063/1.4927434}.
At each time step walkers undergo spawning and death processes, whilst walkers of opposite sign on the same density matrix element annihilate and are removed from the simulation.
Further details of the algorithm can be found elsewhere \citep{Booth2009,PhysRevB.89.245124,:/content/aip/journal/jcp/143/4/10.1063/1.4927434}.

In contrast to path-integral methods, where the sign problem is characterized by an exponential decrease in the average sign with increased system size and decreased temperature, FCIQMC and DMQMC require an exponentially increasing number of walkers for the ground state to emerge from the noise.
The critical number of walkers depends on the annihilation rate, which is much enhanced in a discrete Hilbert space \citep{:/content/aip/journal/jcp/136/5/10.1063/1.3681396,Booth2009,:/content/aip/journal/jcp/138/2/10.1063/1.4773819}.
In practice, the critical population is small enough to allow DMQMC to sample the $N$-particle density matrix \emph{exactly} for system sizes far outside the reach of conventional diagonalization.
The availability of the full density matrix allows arbitrary expectation values to be evaluated without uncontrolled approximations once population-control and time-step biases have been converged \citep{PhysRevB.89.245124,supp}. This applies whether or not the operator of interest commutes with the Hamiltonian.

{\bf \emph{The interaction picture.--}}
It is more efficient to start the simulation from a distribution close to $\hat{\rho}(\beta)$ than from the identity matrix.
With this in mind, we write \mbox{$\hat{H}=\hat{H}^0+\hat{V}$} and consider the quantity \mbox{$\hat{f}(\tau) = e^{-\frac{1}{2}(\beta-\tau) \hat{H}^0} e^{-\tau \hat{H}} e^{-\frac{1}{2}(\beta-\tau)\hat{H}^0}$}, which evolves from \mbox{$e^{-\beta \hat{H}^0}$} at $\tau=0$ to \mbox{$e^{-\beta \hat{H}}$} at $\tau=\beta$
and satisfies a modified Bloch equation,
\begin{equation}
    \frac{d \hat{f}}{d \tau} = \frac{1}{2} \{\hat{H^0}, \hat{f}\} - \frac{1}{2}\Big(\hat{H}_I(-\alpha)\hat{f}+\hat{f}\hat{H}_I(\alpha)\Big) \label{eq:bloch2},
\end{equation}
where \mbox{$\hat{H}_{I}(\alpha) = e^{\alpha \hat{H}^0} \hat{H} e^{-\alpha\hat{H}^0}$} is the interaction-picture Hamiltonian for $\alpha = \frac{1}{2}(\beta-\tau)$. The modified Bloch equation can be simulation in a manner closely analogous to that outlined above. If $\hat{H}^0$ is close to $\hat{H}$, $e^{\frac{1}{2}\tau\hat{H}^0}e^{-\tau\hat{H}}e^{\frac{1}{2}\tau\hat{H}^0}\approx\hat{\mathbb{1}}$ and the statistical fluctuations in estimators are much reduced.

{\bf \emph{The initiator approximation.--}}
It is not possible to sample all of the elements of the density matrix, even using the stochastic algorithm outlined above, but we do not need to as the matrix is extremely sparse.
Rather, we seek a way to find and exploit the sparsity.

We accomplish this by introducing a DMQMC version of the initiator approximation used in FCIQMC simulations \citep{:/content/aip/journal/jcp/132/4/10.1063/1.3302277}. The idea is to restrict the ability of walkers sampling negligibly small density matrix elements to spawn children on other negligibly small matrix elements.
Spawning to already-occupied matrix elements is unaffected, but the initiator approximation only allows spawning events to unoccupied matrix elements if they originate from a set of `initiator determinants' with walker populations above a certain threshold, $n_{\mathrm{add}}$, or if they result from multiple spawning events of the same sign (sign-coherent events) from non-initiator determinants.
The effects of the initiator approximation may be reduced by increasing the total walker population, $N_w$, with the original DMQMC algorithm recovered as $N_w\rightarrow\infty$.
Details of the implementation for DMQMC and a verification that exact results are obtained in the $N_w\rightarrow\infty$ limit can be found in the supplementary material.

{\bf \emph{Basis sets.}--}
For the electron gas we choose the many-particle states to be Slater determinants of plane waves.
The determinantal form builds in the anti-symmetry of the many-particle wavefunction, allowing for an efficient treatment of the exchange processes that dominate as the degeneracy of the system increases.
The dimension of the Hilbert space is restricted by imposing a spherical kinetic energy cutoff $\varepsilon_c = \frac{1}{2} k_c^2$, ensuring that the single-particle basis contains a finite number $M$ of plane waves.
The density matrix is sampled in the corresponding space of $\Big(\sum_{\zeta} {M\choose N_{\uparrow}} {M\choose N_{\downarrow}}\Big)^2$ outer products of Slater determinants, where $\zeta = (N_{\uparrow} - N_{\downarrow})/N$ is the spin polarization, $N_{\uparrow}$ and $N_{\downarrow}$ are the numbers of spin-up and spin-down electrons, and $N = N_{\uparrow} + N_{\downarrow}$. Convergence to the complete basis set ($M\rightarrow\infty$) limit is required to obtain accurate results, a process aided by using the known asymptotic behavior of the internal energy at low and high  temperatures \citep{:/content/aip/journal/jcp/141/16/10.1063/1.4900447,Shepherd2012b,Shepherd2012a,PhysRevB.86.035111,PhysRevB.90.075125,:/content/aip/journal/jcp/143/4/10.1063/1.4927434, supp}.

{\bf \emph{Energies of the warm dense gas.--}}
We are now in a position to provide results for $N=33$, $\zeta=1$, which has emerged as the standard benchmark system for the warm dense electron gas \citep{PhysRevLett.110.146405,PhysRevLett.115.130402,:/content/aip/journal/jcp/143/20/10.1063/1.4936145,PhysRevB.93.085102}.
We focus on the region $0.6\le r_s\le2$ and $0.0625\le \Theta \le 0.5$, where the differences between the restricted and configuration PIMC results are largest and no other data are available \citep{PhysRevLett.115.130402}.
All of the results presented have been carefully checked for convergence with respect to initiator, time step and basis-set errors, and we believe them to be accurate to within the stochastic error bars.
Calculations were performed using the HANDE code \citep{HANDE} with real amplitudes to improve the stochastic efficiency \citep{PhysRevLett.109.230201,:/content/aip/journal/jcp/142/18/10.1063/1.4920975}. More details of the running procedure and precise parameters used can be found in the supplementary material, along with the full $i$-DMQMC data set \citep{supp}.

\begin{figure}
    \includegraphics{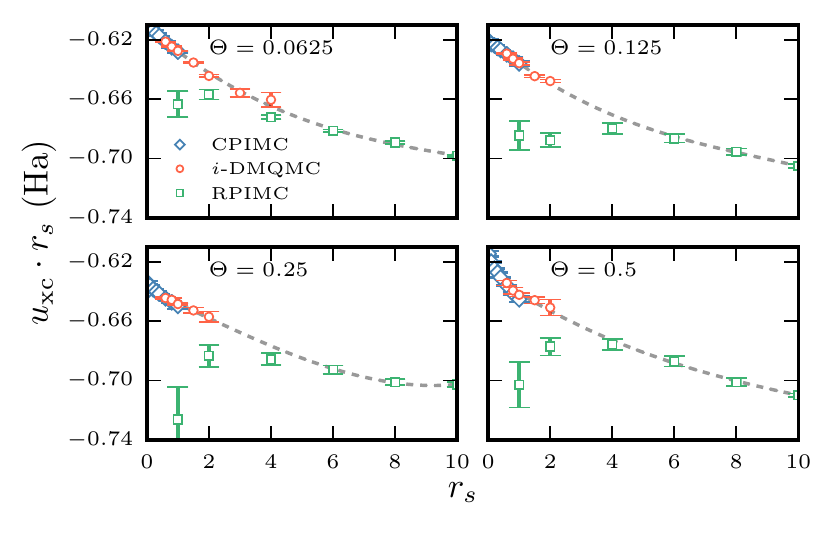}
    \caption{Exchange-correlation energy per particle (times $r_s$) as a function of $r_s$, showing excellent agreement between $i$-DMQMC and configuration PIMC (CPIMC) for $r_s \leq 1$  and differences between $i$-DMQMC and restricted PIMC (RPIMC) for $1\leq r_s\leq 4$.
    For the electron gas, \mbox{$u_{\mathrm{xc}}=(U-U_0)/N$}, where $U_0$ is the internal energy of the $N=33$ non-interacting electron gas in the canonical ensemble.
    Lines are weighted third-order polynomial interpolations \cite{van2011numpy} between the $i$-DMQMC data and the restricted PIMC data for $r_s > 4$ and are meant as guides to the eye.
    The $i$-DMQMC results at $r_s=3$ and $r_s=4$ were obtained using a basis set extrapolation not required at lower $r_s$. The error bars include estimates of the remaining initiator and basis set errors \cite{supp}. \label{fig:exc} }
\end{figure}

The $i$-DMQMC results for the exchange-correlation energy per particle presented in \cref{fig:exc} are in very good agreement with the configuration PIMC results at all values of $r_s$ up to the maximum of $r_s = 1$ considered by Schoof \emph{et al.}\ \citep{PhysRevLett.115.130402}.
(The sign problem prohibited the use of configuration PIMC at higher $r_s$ for the temperatures considered.)
The agreement is even better at lower $r_s$ values.
In particular, our results confirm that the kink-potential approximation used by Schoof \citep{PhysRevLett.115.130402} for $r_s \ge 0.6$ is well controlled and that the restricted PIMC results are significantly too low at $r_s=1$.
Our additional points in the physically important range $1\le r_s \le 2$ ($1\le r_s \le 4$ at low temperatures) further suggest that the restricted PIMC results are unreliable for all $r_s \leq 4$.
We find a slight, apparently systematic, disagreement with configuration PIMC at $\Theta=0.5$, although all points remain within error bars.
The origin of this discrepancy remains unknown but is an active subject of work.

\begin{figure}
    \includegraphics{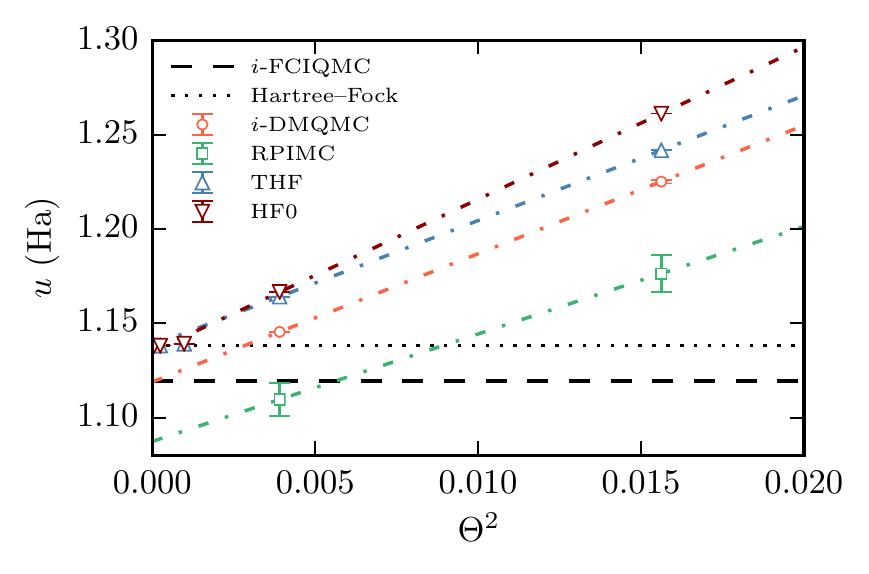}
    \caption{Extrapolation \citep{jones} of the internal energy to $\Theta=0$ for the $N=33$, $\zeta=1$, $r_s=1$ system.
    The restricted PIMC energies are systematically too low and extrapolate to a value considerably below the $i$-FCIQMC ground-state energy.
This discrepancy cannot be explained by finite-size effects.
The $i$-DMQMC (and by extension configuration PIMC) results fare significantly better.
Also shown are two Hartree-Fock-like mean-field estimates (labelled HF0 and THF) of the internal energy in the canonical ensemble \citep{supp}. \label{fig:ground_state_extrap}}.
\end{figure}

As further confirmation of the accuracy of our results, we have carried out independent $i$-FCIQMC calculations \citep{supp} of the internal energy at zero temperature.
Assuming that the energy varies like $T^2$ for small $T$, we can extrapolate the $i$-DMQMC and restricted PIMC results to zero temperature and compare them with the ground state result.
Figure \ref{fig:ground_state_extrap} shows that the extrapolated $i$-DMQMC energy agrees with the ground state result, but that the extrapolated restricted PIMC energy is too low.
This is in contrast to the seemingly reliable extrapolation of the size-corrected restricted PIMC data performed in \citep{PhysRevLett.110.146405}, which agreed well with the Perdew-Zunger parametrization of the local density approximation \citep{PhysRevB.23.5048}.
Also plotted in \cref{fig:ground_state_extrap} are two different finite-temperature mean-field estimates of the internal energy evaluated in the canonical ensemble \citep{supp}, which are seen to perform relatively well.

\begin{figure}
    \includegraphics{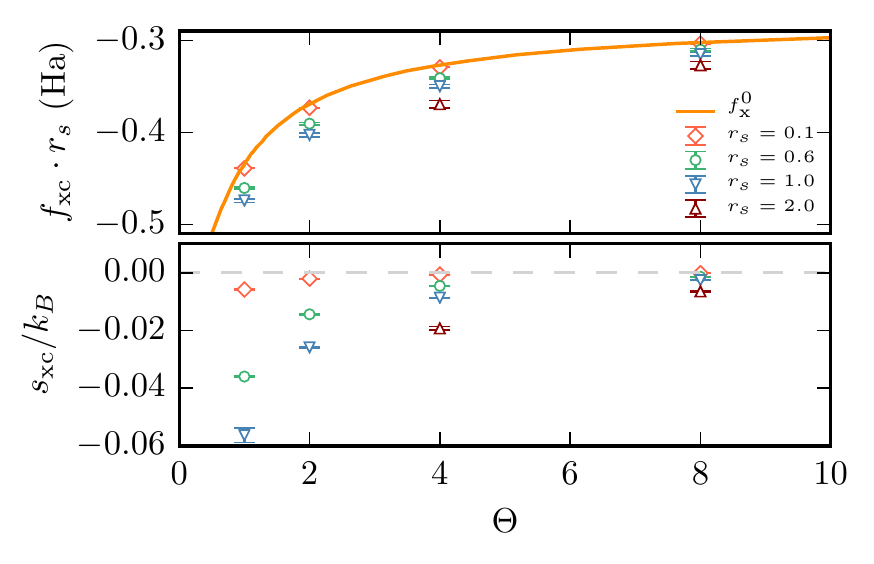}
    \caption{Top panel: Exchange-correlation free energies for the \mbox{$N=33, \zeta=1$} electron gas calculated using DMQMC. The sign problem prohibits calculations below $\Theta=4$ at $r_s=2$. Also plotted is \mbox{$f_\mathrm{x}^0 = N^{-1}\langle \hat{V} \rangle_0$}, the first-order exchange contribution to the free energy evaluated in the canonical ensemble. Bottom panel: The exchange-correlation entropy for the same system. Additional data and more details of the calculation procedures can be found in the supplementary material \cite{supp}.\label{fig:fxc}}
\end{figure}

{\bf \emph{Exact free energies.--}}
The Helmholtz free energy, $F = U - TS = -k_BT \log Z$, where \mbox{$Z = \text{Tr}[\hat\rho]$}, is the quantity required for finite-temperature DFT functionals.
Unfortunately, the entropic term, \mbox{$S=-k_B\mathrm{Tr}[(\hat{\rho}/Z)\log(\hat{\rho}/Z)]$}, requires the logarithm of the density matrix and is difficult to evaluate using traditional QMC methods.
The usual approach is to perform a coupling-constant integration, which, for the electron gas, amounts to an integration of the potential energy over $r_s$ \citep{PhysRevLett.112.076403}.
This requires data at all densities (coupling strengths), increasing the cost of the simulation, and makes use of a possibly unreliable fit to data of unknown functional form.
Attempts to parametrize the finite-temperature exchange-correlation functional require data over the whole $(r_s, \Theta)$ plane, so the additional cost of thermodynamic integration is not an issue. However, for more complicated systems, carrying out a coupling-constant integration may not be possible.
In DMQMC we can evaluate $F_{\mathrm{xc}} = F - F_0$ directly as \citep{supp}
\begin{equation}
    F_{\mathrm{xc}} = k_B T \int_0^{\beta} \langle \hat{V}_I(-\alpha)\rangle_\tau \ d\tau,\label{eq:fxc}
\end{equation}
which is a simple average of the value of a readily available expectation value over the duration of the simulation.
In \cref{fig:fxc} we present data for $f_{\mathrm{xc}} = F_{\mathrm{xc}}/N$ for the \mbox{$N=33,\ \zeta=1$} electron gas evaluated using \cref{eq:fxc}.
We find that $f_{\mathrm{xc}}$ converges slowly with $N_w$ when using the initiator approximation, presumably because of the non-variationality of the form of \cref{eq:fxc}. For this reason, \cref{fig:fxc} is restricted to temperatures $\Theta\ge 1$ where no initiator approximation is required.
More details of this limitation will be explored in a future publication. Also plotted in \cref{fig:fxc} is the exchange-correlation entropy ($s_{\mathrm{xc}} = T^{-1}(u_{\mathrm{xc}}-f_{\mathrm{xc}})$).
As expected, interactions lower the entropy of the system ($S \le S_0$) by an amount that increases with $r_s$ and vanishes in the high $T$ limit.
As $\lim_{T\rightarrow0} s_{\mathrm{xc}} = 0$ and given the behavour of $f_\mathrm{xc}$, we expect $s_{\mathrm{xc}}$ to reach a minimum in the warm dense regime which tends to  counteract a similar minimum found in $u_{\mathrm{xc}}$ \citep{PhysRevLett.110.146405,PhysRevB.93.085102}.
In this regime, $Ts_{\mathrm{xc}}$ is a considerable fraction of $u_{\mathrm{xc}}$ ($\approx 20\%$ at $r_s=\Theta=1$), so the omission of non-ideal entropic effects is expected to be significant.

{\bf \emph{Discussion \& conclusion.--}}
This paper introduced a systematically improvable approximation to DMQMC, allowing for much larger systems to be treated, and used it to study warm dense electron gases with up to $33$ electrons.
Remarkably, even though the largest density matrix sampled has approximately $10^{124}$ elements, we require as few as $10^5$ walkers for certain densities.

Focusing on the canonical test system of $N=33$ spin-polarized particles, we found excellent agreement between \mbox{$i$-DMQMC} and configuration PIMC for $r_s\leq1$ and confirmed that the restricted PIMC results of \citep{PhysRevLett.110.146405} are unreliable at high densities.
In the intermediate to low density regime, we observed significant but decreasing discrepancies persisting up to $r_s\approx4$ and $\Theta \le 0.5$. Our results bridge the gap between the low- and high-density limits and can be used to aid in the parametrization of exchange-correlation functionals for finite-temperature DFT.

Our ground-state calculations confirm that that restricted PIMC internal energies are systematically too low even at low temperatures.
This is inconsistent with the conventional view that the internal energy ought to be variational as $T \rightarrow 0$.
These results are significant because restricted PIMC with free-particle nodes is often considered the most accurate method available to study real warm dense matter systems \citep{PhysRevB.84.224109,PhysRevLett.115.176403}.
Our findings, when combined with the results of configuration and permutation-blocking
PIMC simulations
\citep{PhysRevLett.115.130402,PhysRevB.93.085102,:/content/aip/journal/jcp/143/20/10.1063/1.4936145,PhysRevB.93.205134}, suggest that the free-particle nodal constraint may incur an error of \mbox{5-10\%}, depending on the density and observable considered.
We believe that exponentially scaling, systematically exact methods such as $i$-DMQMC could be of use in analyzing and improving approximations made in restricted PIMC.

The $i$-DMQMC method is complementary to the configuration, direct and restricted PIMC and other novel path-integral \citep{1367-2630-17-7-073017} or finite-temperature FCIQMC approaches \citep{PhysRevLett.115.050603} and is particularly useful at low temperatures, where annihilation and the initiator approximation allow us to overcome the sign problem for surprisingly large systems in many cases.
Open technical challenges remain in the treatment of unpolarized systems and the
development of reliable finite-size corrections at high temperature and density (see,
e.g., the discussion in \citep{PhysRevLett.115.130402} and \citep{PhysRevB.93.205134}), but we are confident that $i$-DMQMC will have an important role to play in the complete characterization of the warm dense electron gas. Finally, given that $i$-DMQMC requires only the Hamiltonian matrix elements, samples the full density matrix, and has direct access to the Helmholtz free energy, it may provide exciting opportunities to investigate the thermodynamics of real warm dense matter exactly.

\emph{Acknowledgements.--}
We thank Tim Schoof, Simon Groth, Tobias Dornheim and Michael Bonitz for helpful discussions facilitated by a CECAM workshop, and for providing unpublished data at early stages of this research.
FDM is funded by an Imperial College PhD Scholarship.
NSB received support from Trinity College, Cambridge and the UK Engineering and Physical Sciences Research Council under grant number EP/J003867/1. JJS thanks the Royal Commission for the Exhibition of 1851 for a Research Fellowship.
FDM, JSS and WMCF acknowledge the research environment provided by the Thomas Young Centre under Grant No.~TYC-101.
WMCF received support from the EPSRC under grant EP/K038141/1.
Computing facilities were provided by the High Performance Computing Service of Imperial College London, by the Swiss National Supercomputing Centre (CSCS) under project ID s523, and by ARCHER, the UK National Supercomputing Service, under EPSRC grant EP/K038141/1 and via a RAP award.

\bibliography{refs}

\end{document}